\documentclass[%
 reprint,
 amsmath,amssymb,
 aps,
]{revtex4-1}

\usepackage{graphicx}
\usepackage{dcolumn}
\usepackage{bm}
\usepackage{float} 

\newcommand*{\affaddr}[1]{#1} 
\newcommand*{\affmark}[1][*]{\textsuperscript{#1}}

\begin{document}

\preprint{APS/123-QED}

\title{Probing Heavy Spin-2 Bosons with $\gamma \gamma$ final states from Vector Boson Fusion Processes at the LHC} 

\author{
Andr\'es Fl\'orez\affmark[2],  Yuhan Guo\affmark[1], Alfredo Gurrola\affmark[1], Will Johns\affmark[1], Oishik Ray\affmark[1], Paul Sheldon\affmark[1], Savanna Starko\affmark[1]\\
\affaddr{\affmark[1] Department of Physics and Astronomy, Vanderbilt University, Nashville, TN, 37235, USA}\\
\affaddr{\affmark[2] Physics Department, Universidad de los Andes, Bogot\'a, Colombia}\\
}

\date{\today}

\begin{abstract}
New massive spin-2 particles are predicted in theoretical extensions to the Standard Model (SM) attempting to solve the hierarchy problem. Such theories postulate that gravity is diluted compared to the other fundamental forces because it can propagate in extra spatial dimensions. While such theoretical models are of high experimental interest because they predict massive spin-2 particles ($Y_{2}$) potentially detectable by collider experiments, searches at the Large Hadron Collider (LHC) have thus far produced no significant evidence for their existence. 
This work considers a hypothetical physics scenario where low coupling strengths between the $Y_{2}$ and quarks/gluons is the underlying reason behind the null $Y_{2}$ search results at the LHC, which have mainly relied on Drell-Yan and gluon-gluon fusion production mechanisms. The focus of this paper is a feasibility study to search for $Y_{2}$ particles using vector boson fusion (VBF) processes at the LHC. In the context of an effective field theory approach with varying couplings $\kappa_{V}$ between $Y_{2}$ and the weak bosons of the SM, we consider the $Y_{2}\to\gamma\gamma$ decay mode to show that the requirement of a diphoton pair combined with two high $p_{T}$ forward jets with large dijet mass and with large separation in pseudorapidity can significantly reduce the SM backgrounds. Assuming proton-proton collisions at $\sqrt{s} = 13$ TeV, we present the total VBF production cross sections, $Y_{2}$ decay widths, and $Y_{2}\to\gamma\gamma$ branching ratios as a function of $m(Y_{2})$, considering universal and non-universal couplings to the SM particles. The unitarity-violating phase space is described. The proposed VBF $Y_{2}\to\gamma\gamma$ search is expected to achieve a discovery reach with signal significance greater than 5$\sigma$ for $Y_{2}$ masses up to 4.4 TeV and $\kappa_{V}$ couplings down to 0.5.
\end{abstract}

\pacs{Valid PACS appear here}
\maketitle


\section{\label{sec:level1}Introduction}

The most important physics result at the CERN Large Hadron Collider (LHC) to date is undoubtedly the observation of a Higgs boson at mass around 125 GeV as reported by the ATLAS and CMS collaborations~\cite{Aad20121,Chatrchyan201230}. With this major discovery and the excellent performance of the LHC, the search for new phenomena at higher energy scales has intensified. While the discovery of new physics can be gleaned from precise measurements of interactions between Standard Model (SM) particles, the issues of the SM also motivate searching for new heavy particles predicted by theoretical extensions addressing its incompleteness. One outstanding issue is the so-called hierarchy problem related with the large difference in scale among the different interactions, i.e. that the electroweak scale is 16 orders of magnitude smaller than the Planck scale at which gravity becomes important. Various mechanisms have been proposed to solve the hierarchy problem. For example, modern theories which attempt to unify gravity with the SM postulate a solution to the hierarchy problem by assuming extra spatial dimensions through which gravity propagates~\cite{Randall:1999ee}. 
While the extent of these theories is vast, the common result is the manifestation of new spin-2 bosons ($Y_{2}$) with TeV scale masses. In the so-called Randall- Sundrum (RS) model, these hypothetical particles are referred to as gravitons.

\begin{figure}[]
 \begin{center} 
 \includegraphics[width=0.5\textwidth]{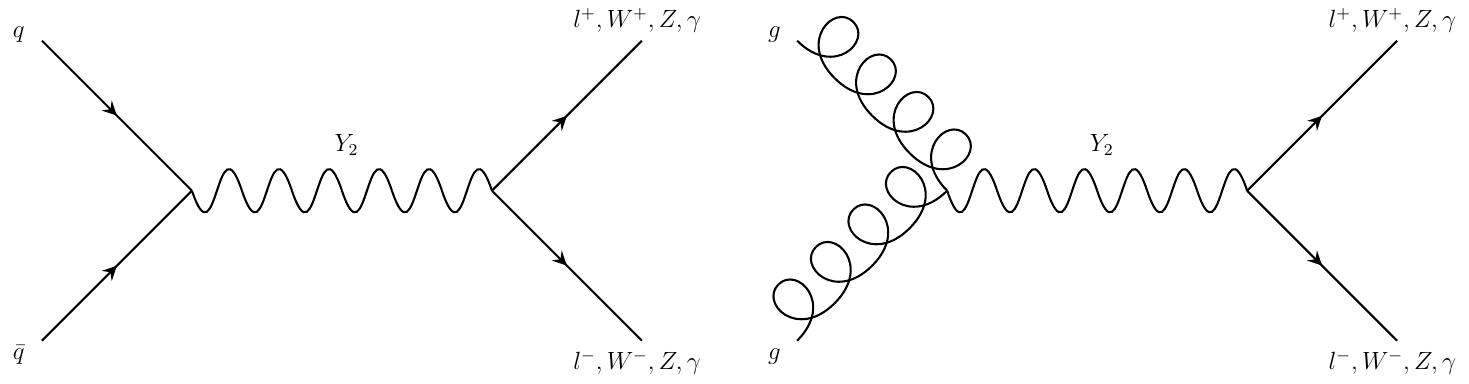}
 \end{center}
 \caption{Representative Feynman diagrams depicting $Y_{2}$ production via quark-antiquark annihilation (left) and gluon-gluon fusion (right).}
 \label{fig:feynDY}
 \end{figure}

Previous LHC searches for spin-2 resonances were performed by the ATLAS~\cite{Aad:2008zzm} and CMS~\cite{Chatrchyan:2008aa} collaborations using proton-proton (pp) collisions at $\sqrt{s} = 8$ and $13$ TeV~\cite{Khachatryan:2016yec1, Aaboud:2015tru, Khachatryan:2016yec2, Aaboud:2016tru, Aaboud:2017tru, Khachatryan:2018VV}. Those results focused on $Y_{2}$ production via gluon-gluon ($gg$) fusion or quark-antiquark ($q\bar{q}$) annihilation (Fig.~\ref{fig:feynDY}), followed by the $Y_{2}$ decay to a pair of photons ($\gamma\gamma$), leptons ($\ell\ell$), jets ($jj$), or heavy vector bosons ($WW$, $ZZ$). Although both collaborations reported the observation of a moderate excess of events compared to SM expectations near diphoton mass $m_{\gamma\gamma} = 750$ GeV with the 8 TeV pp data~\cite{Khachatryan:2016yec1, Aaboud:2015tru}, the 13 TeV results have shown no significant sign of new physics, resulting in exclusion bounds up to $m(Y_{2}) < 4$ TeV depending on the model~\cite{Khachatryan:2016yec2, Aaboud:2016tru, Aaboud:2017tru}. However, exclusions depend strongly on the strength of the coupling between $Y_{2}$ and the quarks/gluons ($\kappa_{q,g}$). While previous searches typically assume more ``natural'' (SM-like) couplings (i.e. $\kappa_{q,g} = 1$), the lack of a strong excess thus far opens up the possibility that $\kappa_{q,g}$ is significantly smaller, in which case the $Y_{2}$ may have eluded discovery. If $\kappa_{q,g}$ is too small to provide large enough cross-section via the traditional $gg\to Y_{2}$ or $q\bar{q}\to Y_{2}$ production mechanisms, another technique must be devised to probe $Y_{2}$.

\begin{figure}[]
 \begin{center} 
 \includegraphics[width=0.5\textwidth]{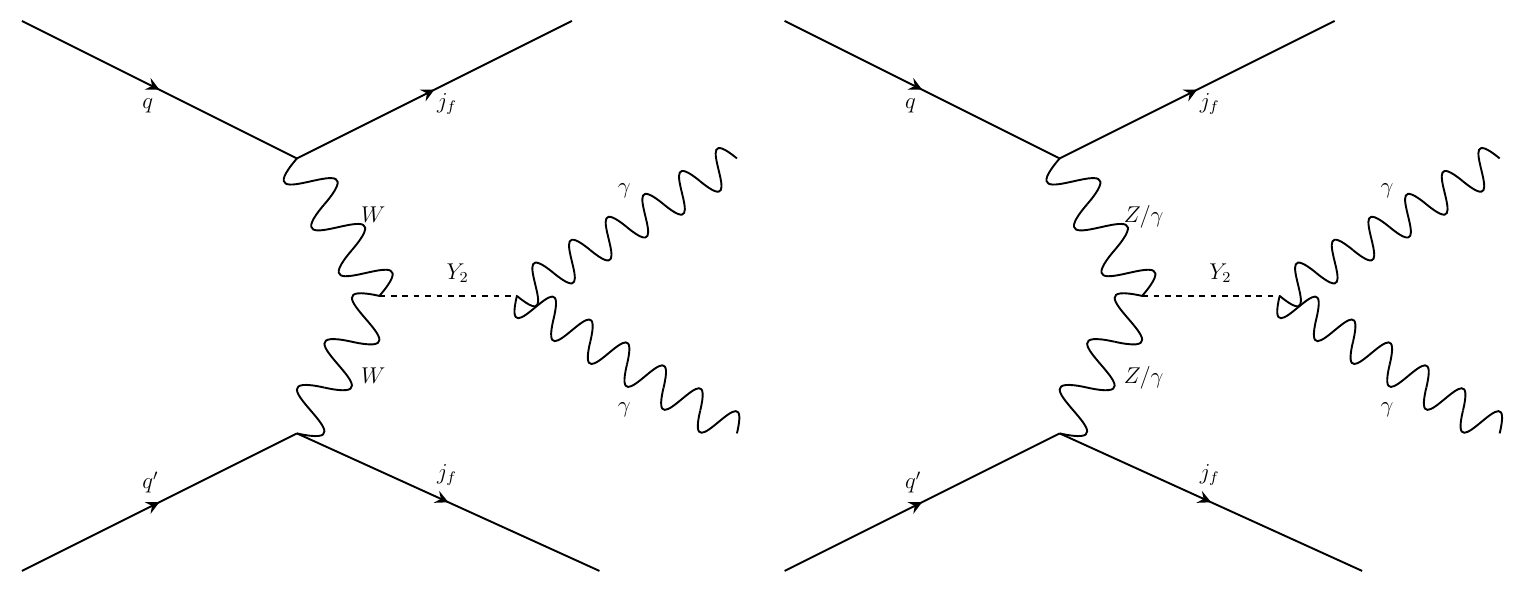}
 \end{center}
 \caption{Representative Feynman diagrams depicting pure electroweak VBF production of a $Y_{2}$ particle and two forward jets. The spin-2 $Y_{2}$ decays to a pair of photons.}
 \label{fig:feynVBF}
 \end{figure}

In this paper, we propose a search for a heavy spin-2 resonance produced through vector boson fusion (VBF) processes and decaying to a pair of photons (Fig.~\ref{fig:feynVBF}). 
The $Y_{2}$ production cross section from VBF processes is independent of $\kappa_{q,g}$. 
The identification of events produced via VBF processes has been a key experimental tool to the discovery of the Higgs boson~\cite{VBFHiggsTauTauCMS}. The VBF topology has also been proposed as an effective tool for dark matter~\cite{DMmodels2, CMSVBFDM}, supersymmetry~\cite{VBF1,VBFSlepton,VBFStop,VBFSbottom,VBF2}, $Z'$~\cite{VBFZprime}, and heavy neutrino searches at the LHC~\cite{VBFHN} due to the natural collinear logarithmic enhancement obtained in the production cross section at high mass values, as well as the significant reduction of SM backgrounds.  The $Y_{2}\to \gamma \gamma$ decay channel is motivated by its relatively clean experimental signature with localized diphoton mass spectrum. Furthermore, while spin-1 $Z'$ and spin-2 $Y_{2}$ bosons can both result in dilepton, dijet, and $ZZ$/$WW$ channels, an excess in the $\gamma \gamma$ final state would rule out a $Z'$.

To allow for a generic and model independent approach, we consider an effective field theory (EFT) consisting of three terms in the Lagrangian: ($i$) kinetic term $\mathcal{L}_{V,f} = -\frac{\kappa_{V,f}}{\Lambda}T_{\mu\nu}^{V,f}Y_{2}^{\mu\nu}$ representing the $Y_{2}$ interactions with the SM gauge (matter) fields $V$ ($f$) and where $T_{\mu\nu}^{V,f}$ is the energy-momentum tensor of $V$ or $f$; ($ii$) $Y_{2}$ interaction with the SM Higgs doublet $\Phi$, $\mathcal{L}_{\Phi} = -\frac{\kappa_{H}}{\Lambda}T_{\mu\nu}^{\Phi}Y_{2}^{\mu\nu}$ with $\Phi$ energy-momentum tensor $T_{\mu\nu}^{\Phi}$; and ($iii$) $Y_{2}$ interactions with Fadeev-Popov (FP) ghost fields $\mathcal{L}_{\textrm{FP}} = -\frac{\kappa_{V}}{\Lambda}T_{\mu\nu}^{\textrm{FP}}Y_{2}^{\mu\nu}$, where $T_{\mu\nu}^{\textrm{FP}}$ is the energy-momentum tensor of the FP ghost field. In the above Lagrangian equations $\Lambda$ represents the cut-off scale, the gauge fields $V$ are the SM electroweak gauge bosons or the gluon, and the matter fields $f$ are quarks, leptons, and left-handed neutrinos. This particular implementation also allows for scenarios with non-universal couplings to be studied at NLO accuracy~\cite{EFTSpin2}.

\section{Samples and simulation}

Signal and background samples were generated with MadGraph5\_aMC (v2.6.4)~\cite{MADGRAPH}, considering proton-proton (pp) beams with $\sqrt{s}=13$ TeV. 
In the case of the spin-2 signal samples, the model files were generated using the FeynRules package~\cite{FeynRules} and obtained from Ref.~\cite{spin2UFO}. The dominant contributions to the total $Y_{2}$ decay width are from the massive SM final states (i.e. $Y_{2}\to t\bar{t}$, $Y_{2}\to W^{+}W^{-}$, $Y_{2}\to ZZ$, and $Y_{2}\to HH$), with a dependence given by $\Gamma(Y_{2}\to XX) \approx \kappa_{X}^{2}\frac{m(Y_{2})^{3}}{\Lambda^{2}}(1 - 4\frac{m(X)^{2}}{m(Y_{2})^{2}})^{(2l+1)/2}$, where $l$ is the angular momentum quantum number, $m(X)$ is the mass of the heavy particle $X$, and $\kappa_{X}$ the coupling of $X$ to $Y_{2}$. The term $(1 - 4\frac{m(X)^{2}}{m(Y_{2})^{2}})^{(2l+1)/2}$ represents the kinematically allowed phase space, i.e. $4\frac{m(X)^{2}}{m(Y_{2})^{2}} < 1$ and thus $m(Y_{2}) > 2m(X)$. Therefore, since we allow the possibility that $Y_{2}$ couples to all the heavy SM particles, we consider $m(Y_{2})$ values above $2m(t) \approx 350$ GeV.

\begin{figure}[]
 \begin{center} 
 \includegraphics[width=0.45\textwidth]{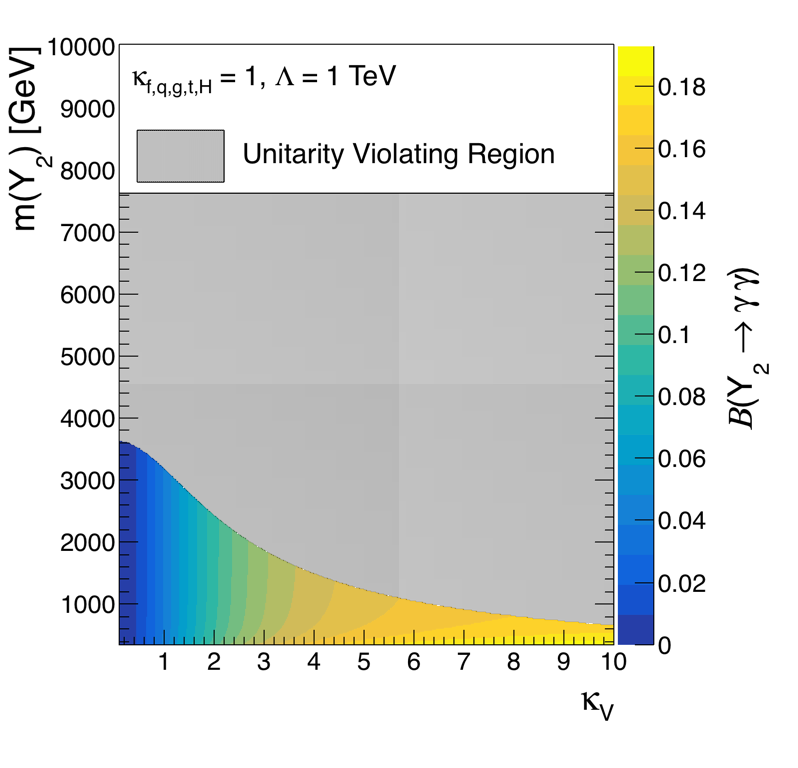}
 \end{center}
 \caption{$Y_{2}\to \gamma \gamma$ branching ratio as a function of $m(Y_{2})$ and $\kappa_{V}$, for the benchmark case where $\kappa_{f,g,H}=1$. The gray shaded region represents the unitarity-violating phase space.}
 \label{fig:UVphasespaceA}
 \end{figure}

\begin{figure}[]
 \begin{center} 
 \includegraphics[width=0.45\textwidth]{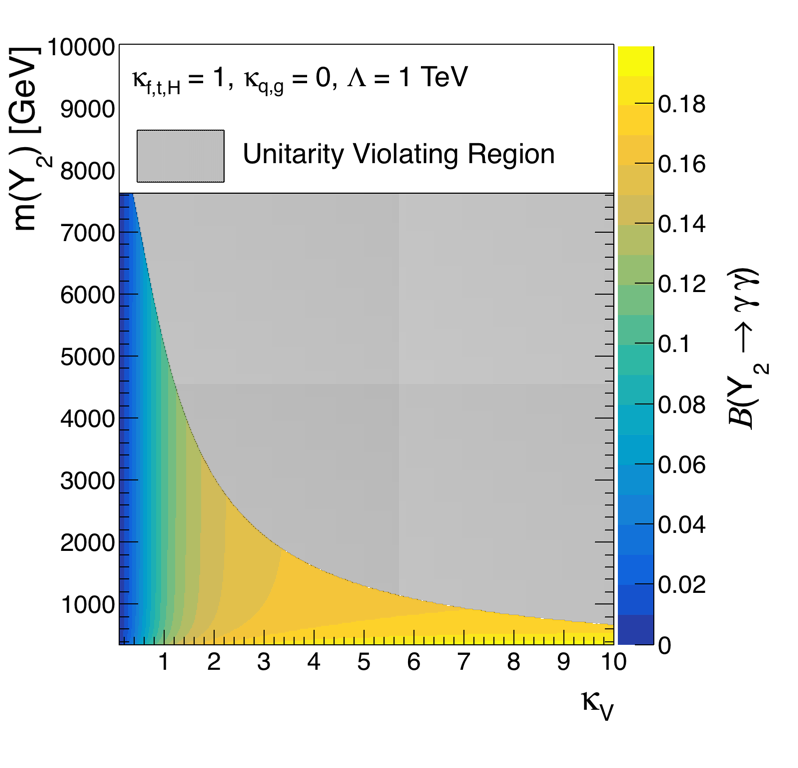}
 \end{center}
 \caption{$Y_{2}\to \gamma \gamma$ branching ratio as a function of $m(Y_{2})$ and $\kappa_{V}$, for the benchmark case where $\kappa_{q,g}=0$. The gray shaded region represents the unitarity-violating phase space.}
 \label{fig:UVphasespaceB}
 \end{figure}

To illustrate the unitarity violating phase space, we consider two benchmark scenarios: $(i)$ standard couplings to the fermions, gluons, and Higgs boson ($\frac{\kappa_{f,g,H}}{\Lambda} = 1$ TeV$^{-1}$), but with a varying coupling $\kappa_{V} = \kappa_{\gamma,W,Z}$ to the electroweak bosons; and $(ii)$ suppressed couplings to the gluon and light quarks $q = u,d,s,c,b$ ($\frac{\kappa_{q,g}}{\Lambda} = 0$), but with free parameter $\kappa_{V}$. The latter case represents a proxy for our main interest in this paper, where the lack of a strong excess in $gg\to Y_{2}$ and $q\bar{q}\to Y_{2}$ searches thus far motivates the possibility that $\kappa_{q,g}$ is small. This scenario makes VBF an important mode for discovery. 
On the other hand, it is important to point out that even if $\kappa_{q,g} \approx 1$ (or larger), a VBF $Y_{2}$ search is still relevant to establish the coupling of $Y_{2}$ to the vector bosons of the SM, which is important to assess the correct physics model that best fits the observed data should there be evidence for discovery. Fig.~\ref{fig:UVphasespaceA} shows the unitarity-violating phase space and the $Y_{2}\to \gamma \gamma$ branching ratio, as a function of 
$m(Y_{2})$ and $\kappa_{V}$, for the benchmark case where we assume $\kappa_{f,g,H}=1$. Fig.~\ref{fig:UVphasespaceB} shows the unitarity-violating phase space and the $Y_{2} \to \gamma \gamma$ branching ratio, as a function of 
$m(Y_{2})$ and $\kappa_{V}$, for the benchmark case where we assume $\kappa_{q,g}=0$. In the scenario with more natural couplings to the light quarks and gluon, the unitarity acceptable region extends to $m(Y_{2}) \approx$ 3.6 (3.2) TeV for $\kappa_{V} =$ 0.1 (1.0). The theoretically allowed region can be as large as $m(Y_{2}) \approx$ 8.4 (5.0) TeV for $\kappa_{V} =$ 0.1 (1.0) when $\kappa_{q,g} = 0$. For this reason we generate signal samples with $m(Y_{2})$ values up to 8.4 TeV. The diphoton branching ratio $B(Y_{2} \to \gamma \gamma)$ is approximately 10 (4)\% for $\kappa_{q,g}=0$ (1) and $\kappa_{V} =$ 1.0, but can be as high as about 20\% for large $\kappa_{V}$ values. 

\begin{figure}[]
 \begin{center} 
 \includegraphics[width=0.45\textwidth]{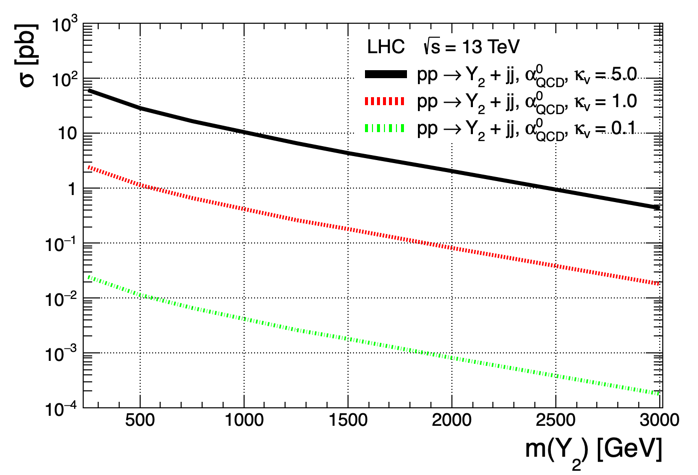}
 \end{center}
 \caption{The VBF $Y_{2}$ cross-section as a function of $m(Y_{2})$ and $\kappa_{V}$.}
 \label{fig:XSections}
 \end{figure}

\begin{figure}[]
 \begin{center} 
 \includegraphics[width=0.45\textwidth]{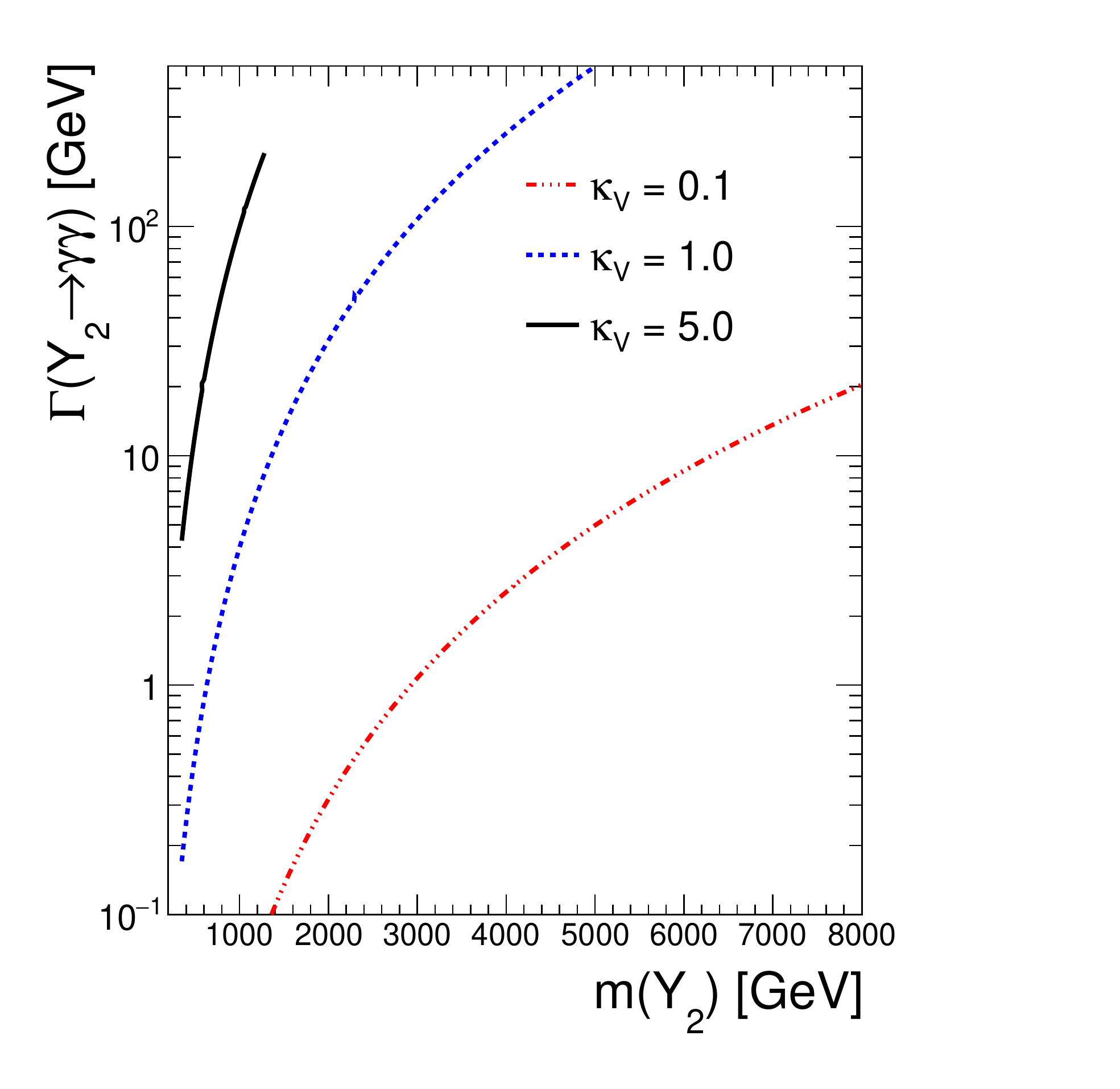}
 \end{center}
 \caption{The $Y_{2}\to\gamma\gamma$ decay width as a function of $m(Y_{2})$ and $\kappa_{V}$.}
 \label{fig:DecayWidth}
 \end{figure}

The signal samples were produced considering electroweak production of $Y_{2}$ with two associated jets and suppressed QCD coupling (i.e. $pp\to jjY_{2}$ with $\alpha_{QCD}^{0}$). Fig.~\ref{fig:XSections} shows the VBF $Y_{2}$ production cross section as a function of $m(Y_{2})$ for $\kappa_{V} =$ 0.1, 1.0, and 5.0. As expected, the cross section scales as $\kappa_{V}^{2}$. Fig.~\ref{fig:DecayWidth} shows the $Y_{2}\to\gamma\gamma$ decay width as a function of $m(Y_{2})$ for the three similar values of $\kappa_{V}$. The curves in Fig.~\ref{fig:DecayWidth} do not extend into the unitarity violating phase space.

The dominant sources of SM background are production of photon pairs with associated jets, referred to as $\gamma\gamma$+jets. In the proposed search region (defined in Section III), the associated jets are mainly from initial state radiation (i.e. $pp \to \gamma \gamma jj$, $\alpha_{QCD}^{2}$) or SM VBF processes (i.e. $pp \to \gamma \gamma jj$, $\alpha_{QCD}^{0}$). Therefore, the background samples are split into two categories: $(i)$ non-VBF $\gamma\gamma$+jets events with up to four associated jets, inclusive 
in the electroweak coupling ($\alpha_{EWK}$) and $\alpha_{QCD}$; and $(ii)$ pure electroweak VBF $\gamma\gamma jj$. The production of $\gamma$+jets and multijet events with jets misidentified as photons have been checked to provide a negligible contribution to the proposed search region due to the effectiveness of the VBF selection criteria. PYTHIA (v6.416)~\cite{Sjostrand:2006za} was used for the hadronization process in both signal and background samples, while Delphes  (v3.3.2)~\cite{deFavereau:2013fsa} was used to simulate detector effects using the CMS parameters. At MadGraph level, photons were required to have transverse momentum greater than 10 GeV and located in the central region of the ATLAS and CMS detectors ($|\eta (\gamma)| < 2.5$). Photon pairs were also required to be separated in $\eta$-$\phi$ space by requiring $\Delta R(\gamma_{1},\gamma_{2}) = \sqrt{\Delta \phi(\gamma_{1}, \gamma_{2})^{2}+\Delta \eta(\gamma_{1}, \gamma_{2})^{2}} >0.4$. Similarly, at parton level the jets were required to have a minimum $p_{T}> 20$ GeV and $|\eta| < 5.0$. 
The MLM algorithm \cite{MLM} was used for jet matching and jet merging. 
The xqcut and qcut variables of the MLM algorithm, related with the minimal distance between partons and the energy spread of the clustered jets, were set to 30 and 45 as result of an optimization process requiring the continuity of the differential jet rate as a function of jet multiplicity. 

\section{Event selection criteria}

Although VBF processes are characterized by relatively low production cross sections (compared to DY and gluon-gluon fusion processes), its unique topology compensates for the smaller production rate by helping to significantly suppress SM backgrounds, especially those with jets arising from QCD interactions. 
The VBF topology consists of two high-$p_{T}$ forward jets, in opposite sides of the detector, with a large difference in pseudorapidity and TeV scale dijet mass.

\begin{table}[]
\begin{center}
\caption {Event selection criteria.}
\label{tab:selection_criteria}
\begin{tabular}{ l  c c}\hline\hline
Criterion & $\gamma_1\gamma_2 j_{f}j_{f}$\\
 \hline
  \multicolumn{3}{ c }{{\bf Central Selections}} \\
   \hline
   $|\eta(\gamma)|,|\eta(e)|,|\eta(\mu)|,|\eta(\tau_{h})|$ &  & $< 2.5$\\ $|\eta($b$)|$ &  & $< 2.4$\\
   $p_{T}(\gamma)$ & & $> 30$ GeV\\
   $p_{T}(e),p_{T}(\mu)$ & & $> 10$ GeV\\
   $p_{T}(\tau_{h}),p_{T}($b$)$ & & $> 20$ GeV\\
   $N(\gamma)$ & & $= 2$\\
   $N(e),N(\mu),N(\tau_{h}),N($b-jets$)$ & & $= 0$\\
   $p_{T}^{\textrm{lead}}(\gamma)$ & & $> 60$ GeV\\
   $\Delta R(\gamma_{1},\gamma_{2})$ & & $> 0.4$\\ 
   \hline
   \multicolumn{3}{ c }{{\bf VBF Selections}} \\
   \hline
   $p_{T}(j)$ & & $>30$ GeV\\
   $|\eta(j)|$ & & $< 5.0$\\
   $\Delta R(\gamma, j)$ & & $> 0.4$\\
   $N(j)$ & & $\geq$ 2\\
   $\eta(j_{1})\cdot \eta(j_{2})$ & & $< 0$\\
   $|\Delta \eta (j_{1}, j_{2})| $ & & $> 4.0$ (``Loose'') \\
    & & $> 6.0$ (``Tight'')\\
   $m_{jj}$ & & $> 1.0$ TeV \\
   \hline\hline
 \end{tabular}
\end{center}
\end{table}

\begin{figure}[]
    \centering
    \includegraphics[width=0.65\textwidth]{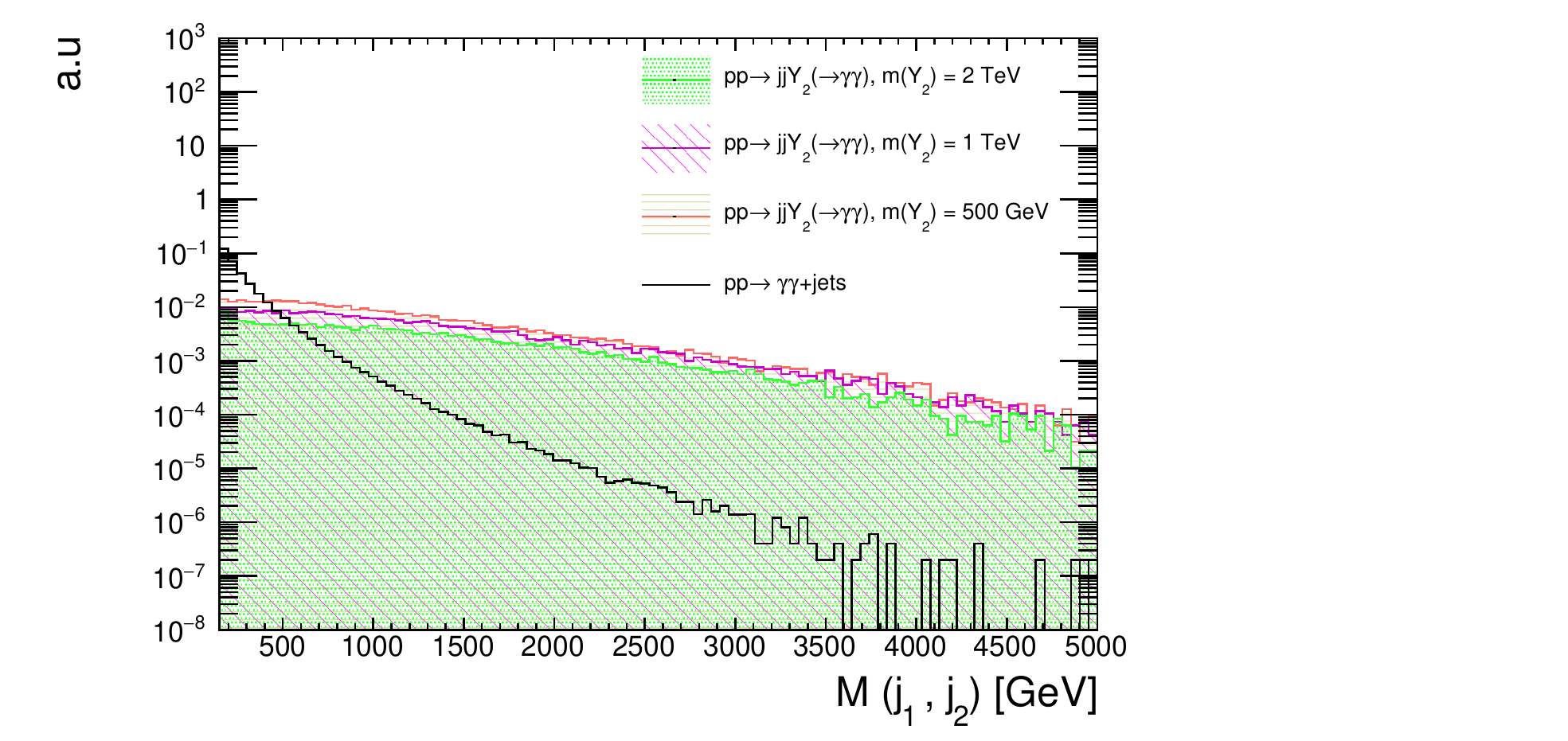}
    \caption{The dijet mass distribution (normalized to unity) for the total SM backgrounds and $m(Y_2) = 2$ TeV signal benchmark point.}
    \label{fig:mjj_distr}
\end{figure}

In the higher energy LHC environment, where $m(j) \ll p_{T}(j)$ and $E_{T} \approx p_{T}$, the dijet mass $m_{jj}$ is well-approximated by $m_{jj} \approx \sqrt{2p_{T}^{j_{1}}p_{T}^{j_{2}}\text{cosh}(\Delta\eta_{jj})}$.
The transverse momentum requirement on the jets is driven by the experimental constraints of the CMS and ATLAS experiments. These constraints include the limitations of the reconstruction algorithms and the geometric acceptance of the detectors. Events are required to have at least two jets with $p_{T}(j) > 30$ GeV and $|\eta(j)| < 5.0$.
Since $m_{jj}$ depends on the $p_T$ of each jet, we indirectly optimize the $p_T$ values by optimizing $m_{jj}$.
Considering the large pseudorapidity gap $|\Delta\eta_{jj}|$ in VBF $Y_{2}$ events, we impose a stringent requirement on the lower threshold of the calculated dijet mass. 

Shown in Fig.~\ref{fig:mjj_distr} are the $m_{jj}$ distributions (normalized to unity) of the total background and the signal benchmark point with $m(Y_2) = 2$ TeV. The signal distribution is broad and overtakes the SM backgrounds at TeV scale $m_{jj}$.
The exact $m_{jj}$ requirement, $m_{jj} > 1$ TeV, is determined through an optimization process that maximizes signal significance. For this purpose we take a simple approach to defining signal significance as $N_{\textrm{S}} / \sqrt{N_{\textrm{S}} + N_{\textrm{B}} + (0.25N_{\textrm{B}})^{2}}$, where $N_{\textrm{S}}$ and $N_{\textrm{B}}$ are the signal and background yields, respectively, while the factor of $0.25N_{\textrm{B}}$ represents a 25\% systematic uncertainty on the background (to be discussed later). The values of $N_{\textrm{S}}$ and $N_{\textrm{B}}$ in the optimization process are derived within a diphoton mass window $m(Y_{2}) - 2\Gamma < m_{\gamma\gamma} < m(Y_{2}) + 2\Gamma$, where $\Gamma$ is the $Y_{2}\to\gamma\gamma$ decay width. We note these particular definitions of $N_{\textrm{S}}$, $N_{\textrm{B}}$, and significance are used only for the purpose of optimizing the selections. We propose to determine final discovery potential with a shape based analysis to be described in Section~\ref{sec:Results}.

\begin{figure}[]
    \centering
    \includegraphics[width=0.5\textwidth]{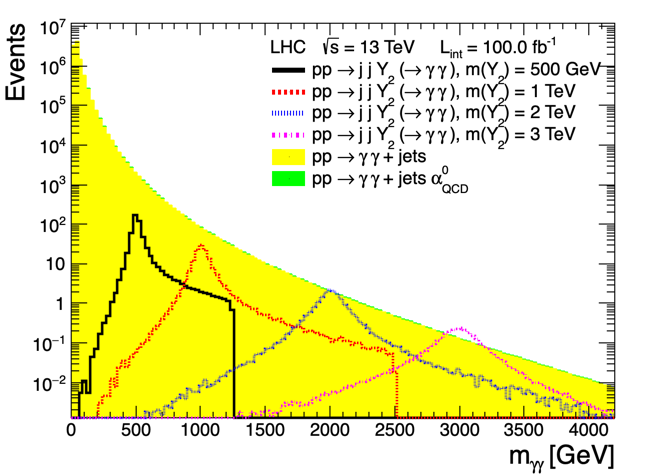}
    \caption{$m_{\gamma\gamma}$ distributions for the main backgrounds and various signal benchmark points. Simulated events populating these distributions have satisfied the requirements of at least two jets and two photons. A comparison of the signal and background distributions indicate the need to apply more stringent VBF selections for the low-mass signal benchmark points than the high-mass scenarios.}
    \label{fig:maa_no_cut}
\end{figure}

\begin{figure}[]
    \centering
    \includegraphics[width=0.5\textwidth]{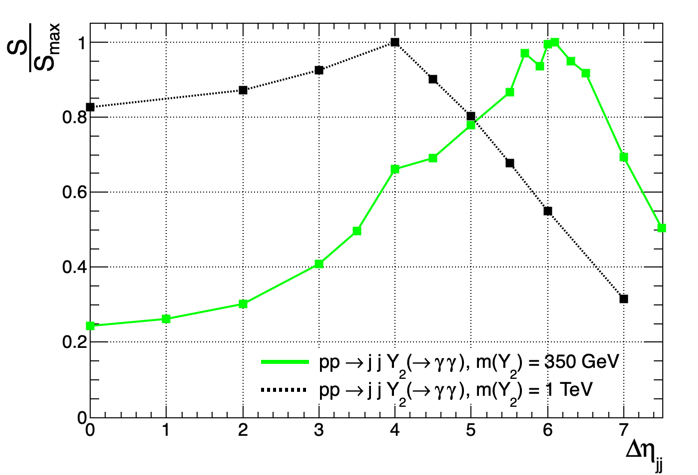}
    \caption{
    The normalized signal significance $\frac{S}{S_\text{max}}$ as a function of the lower bound imposed on $|\Delta\eta_{jj}|$.
    The critical point at 6.0 for $m(Y_2) = $ 350 GeV motives us to apply a ``Tight'' selection cut of $|\Delta\eta| > 6.0$ to maximize discovery potential for low mass signal models. Similarly, the critical point at 4.0 for $m(Y_2) = $ 1 TeV motives us to apply a ``Loose'' selection cut of $|\Delta\eta| > 4.0$ to maximize discovery potential for high mass (TeV scale) signal models.}
    \label{fig:delta_eta_sig_norm}
\end{figure}

The optimal selection cut for the pseudorapidity gap $|\Delta\eta_{jj}|$ depends on $m (Y_2)$. 
Fig.~\ref{fig:maa_no_cut} shows the distribution of the reconstructed diphoton mass ($m_{\gamma\gamma}$) for the main backgrounds and various signal benchmark points. 
The contribution from SM backgrounds decreases exponentially as $m_{\gamma\gamma}$ increases.
On the other hand, the $m_{\gamma\gamma}$ signal distribution appears as a localized bump centered at approximately $m(Y_{2})$.
Since the signal benchmark points with small $m(Y_2)$ have low values of reconstructed $m_{\gamma\gamma}$ lying within the bulk of the background distributions, a more stringent $|\Delta\eta_{jj}|$ cut is required to efficiently reduce the SM backgrounds. 
The signal benchmark points with high $m (Y_2)$ values, on the other hand, lie in the tail of the reconstructed $m_{\gamma\gamma}$ background distribution, thus requiring a less stringent $|\Delta\eta_{jj}|$ cut.

Fig.~\ref{fig:delta_eta_sig_norm} shows the normalized signal significance $S/S_\text{max}$ as a function of $|\Delta\eta_{jj}|$ cut value, assuming an integrated luminosity of 100 fb$^{-1}$. The normalized significance is 
the significance at a given cut value, divided by the maximum significance over all possible cut values of the variable under study. 
The signal significance for $m(Y_{2}) = 350$ GeV is maximized with $|\Delta\eta_{jj}| > 6.0$, while $|\Delta\eta_{jj}| > 4.0$ provides the best discovery potential for the higher mass signal point $m(Y_{2}) = 1$ TeV. 
We hence define a ``Tight'' and ``Loose'' signal region with pseudorapidity gap requirements of $|\Delta\eta_{jj}| > 6.0$ and $|\Delta\eta_{jj}| > 4.0$, respectively, in an attempt to maximize the sensitivity to low and high mass signatures. 
The results shown in Section~\ref{sec:Results} are obtained by applying both ``Tight'' and ``Loose'' criteria and choosing the one that gives the largest signal significance for a given $m(Y_{2})$ value.

\begin{figure}[]
    \centering
    \includegraphics[width=0.65\textwidth]{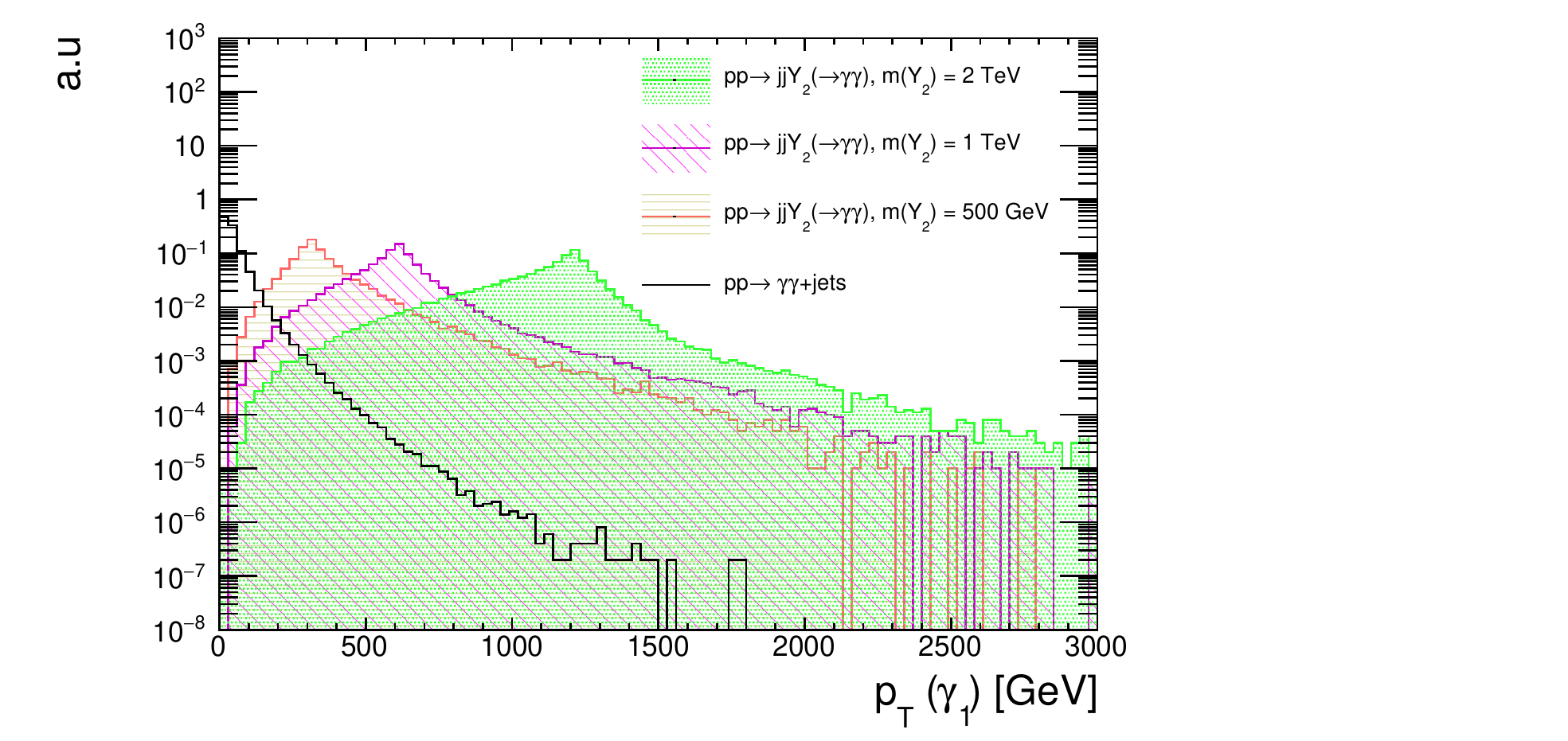}
    \caption{$p^\text{lead}_{T}(\gamma)$ distributions, normalized to unity, for the main backgrounds and various signal benchmark points.}
    \label{fig:pt_a1}
\end{figure}

\begin{figure}[]
    \centering
    \includegraphics[width=0.5\textwidth]{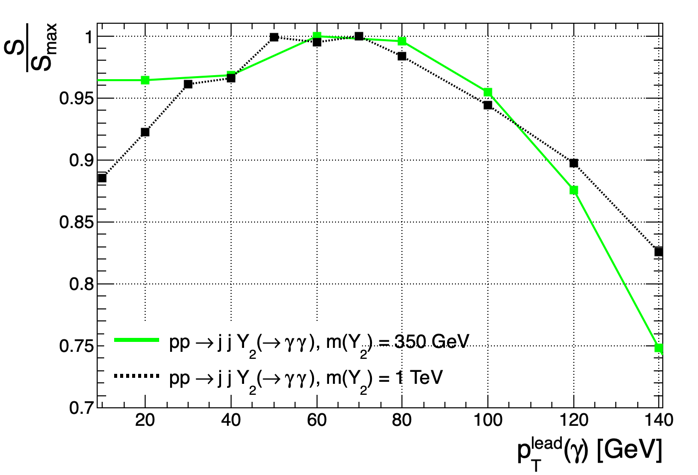}
    \caption{The normalized signal significance $\frac{S}{S_\text{max}}$ as a function of the lower bound imposed on $p^\text{lead}_{T}(\gamma)$.
    The signal significance is maximized at around around 60 GeV for both benchmark points with $m(Y_2) = 350$ GeV and $m(Y_2) = 1$ TeV, which motives a selection cut at $p_T^\text{lead}(\gamma) > 60$ GeV.}
    \label{fig:pt_a1_sig_norm}
\end{figure}

In addition to the VBF selections, we select two $\gamma$ candidates, requiring that $p_{T}(\gamma) > 30$ GeV and $|\eta(\gamma)| < 2.5$ for both photons with $p_T^\text{lead}(\gamma) > 60$ GeV for the leading photon.
The signal and background $p_{T}$ distributions (normalized to unity) of the leading photon are shown in Fig.~\ref{fig:pt_a1}, while Fig.~\ref{fig:pt_a1_sig_norm} shows the normalized signal significance $S/S_\text{max}$ as a function of $p^\text{lead}_{T}(\gamma)$ cut value, assuming an integrated luminosity of 100 fb$^{-1}$.
The signal significance is maximized at approximately 60 GeV for both the $m(Y_2) = 350$ GeV and the $m(Y_2) = 1$ TeV benchmark points, which supports our choice of selection cut at $p_T^\text{lead}(\gamma) > 60$ GeV. 
By a similar methodology, we determined the optimal threshold for the $p_{T}$ of the sub-leading photon to be $p_{T}(\gamma)> 30$ GeV.

Finally, in addition to the VBF dijet and diphoton selections, simulated events in the proposed search region are required to have zero jets tagged as b-quarks and no reconstructed/identified leptons (electrons, muons, or hadronically decay taus). While these selections are highly efficient for VBF $Y_{2}\to\gamma\gamma$ signal events ($> 90$\%), they simultaneously help to ensure the negligible  contribution from SM backgrounds with top quarks, pairs of vector bosons, $Z/\gamma*\to\ell\ell$ with associated jets, and $W\to\ell\nu$ with associated jets. Electrons and muons used in the lepton veto requirement must have $p_{T} > 10$ GeV and $|\eta| < 2.5$, while hadronically decaying tau leptons ($\tau_{h}$) have $p_{T} > 20$ GeV and $|\eta| < 2.5$. Similarly, b-jets are defined by $p_{T} > 20$ GeV and $|\eta| < 2.4$. The difference in the $\tau_{h}$/b-jet $p_{T}$ thresholds, in comparison to the electron/muon thresholds, is due to the experimentally more challenging nature of $\tau_{h}$ and b-jet reconstruction. The event selection criteria used in the $\gamma_{1}\gamma_{2} j_{f}j_{f}$ ``Tight'' and ``Loose'' signal regions are summarized in Table~\ref{tab:selection_criteria}.

\begin{figure}[]
    \centering
    \includegraphics[width=0.5\textwidth]{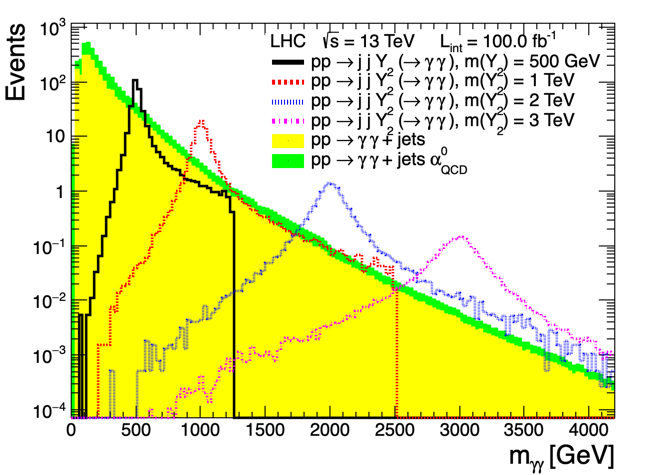}
    \caption{$m_{\gamma\gamma}$ distribution for the main backgrounds and various signal benchmark points, after applying the event selection criteria as in Table~\ref{tab:selection_criteria} with $|\Delta\eta| > 4.0$.}
    \label{fig:maa_after_cuts_loose}
\end{figure}
\begin{figure}[]
    \centering
    \includegraphics[width=0.5\textwidth]{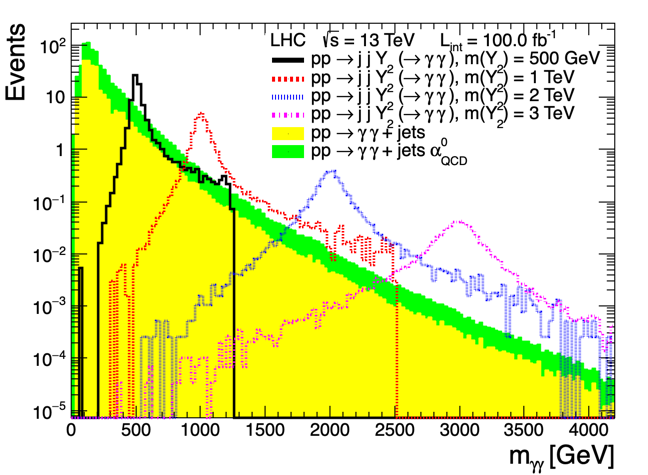}
    \caption{$m_{\gamma\gamma}$ distribution for the main backgrounds and various signal benchmark points, after applying the event selection criteria as in Table~\ref{tab:selection_criteria} with $|\Delta\eta| > 6.0$.}
    \label{fig:maa_after_cuts_tight}
\end{figure}
Fig.~\ref{fig:maa_after_cuts_loose} shows the signal and background $m_{\gamma\gamma}$ distributions, after applying the ``Loose'' event selection criteria as listed in Table~\ref{tab:selection_criteria} with $|\Delta\eta| > 4.0$. Fig.~\ref{fig:maa_after_cuts_tight} is similar but with the ``Tight'' $|\Delta\eta|$ requirement of 6.0. Again, the bulk of the (now largely reduced) background distribution lies at relatively low reconstructed diphoton mass values, while the signal bumps dominate at the tails of the $m_{\gamma\gamma}$ distribution. 
Depending on the $\Delta\eta$ cut value, the contributions of the $\gamma\gamma$+jets backgrounds are reduced by approximately 4-5 orders of magnitude, while the signal acceptance is 12-58\% depending on $m(Y_{2})$.

\section{Results}\label{sec:Results}

We use the full range of the reconstructed diphoton mass distribution to perform a shape based analysis of the discovery potential, following a profile binned likelihood approach to define the test statistic, using the ROOFit \cite{ROOTFit} toolkit. Systematic uncertainties are incorporated into the calculation as nuisance parameters in order to obtain an experimentally realistic estimate of the signal significance. 
A 15\% systematic uncertainty related with $\gamma$ identification efficiency, based on \cite{Khachatryan:2016yec1,Aaboud:2015tru}, has been included, considering 100\% correlation between the two photons and 100\% correlation among signal and background processes. In addition, a 12\% uncertainty in the signal and background yields due to the choice of parton distribution function (PDF) included in the simulated signal samples is evaluated based on the PDF4LHC recommendations~\cite{Butterworth:2015oua}. This is consistent with the 15\% uncertainty from the CMS VBF SUSY and VBF dark matter searches~\cite{VBF2,CMSVBFDM}. The effect of the chosen PDF set on the shape of the  $m_{\gamma\gamma}$ distribution is negligible. Finally, we have included a 20\% uncertainty related with the experimental difficulties involved with reconstructing, identifying, and calibrating forward jets at the LHC. This 20\% effect is categorized as an overall uncertainty on the efficiency of the VBF selection criteria, based on Refs.~\cite{VBF2,CMSVBFDM}. 
A local p-value is calculated as the probability under a background only hypothesis to obtain a value of the test statistic as large as that obtained with a signal plus background hypothesis. 
The signal significance $S$ is then defined as the value at which the integral of a Gaussian between $S$ and $\infty$ results in a value equal to the local p-value.

\begin{figure}[]
 \begin{center} 
 \includegraphics[width=0.5\textwidth, height=0.35\textheight]{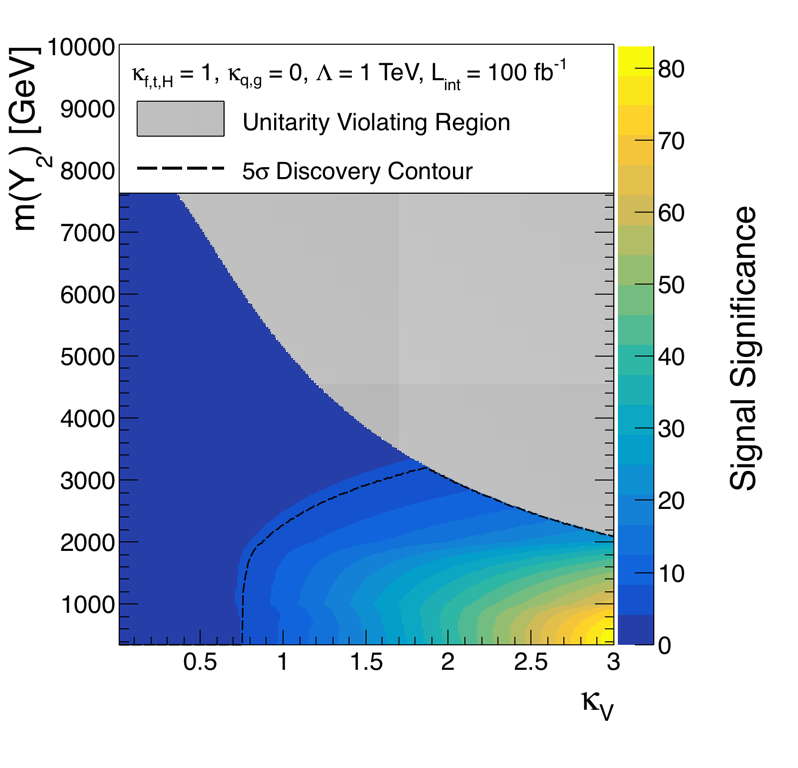}
 \end{center}
 \caption{Expected signal significance for the proposed VBF final state. The results are shown as $m(Y_{2})$ vs. $\kappa_{V}$ on the $y-x$ plane, and the expected signal significance on the $z-$axis.  The expected signal significance was calculated performing a profile binned likelihood of the diphoton mass distribution, assuming an expected luminosity of 100 $fb^{-1}$. The black dashed line encloses the region with 5$\sigma$ discovery potential.}
 \label{fig:results}
 \end{figure}

\begin{figure}[]
 \begin{center} 
 \includegraphics[width=0.5\textwidth, height=0.35\textheight]{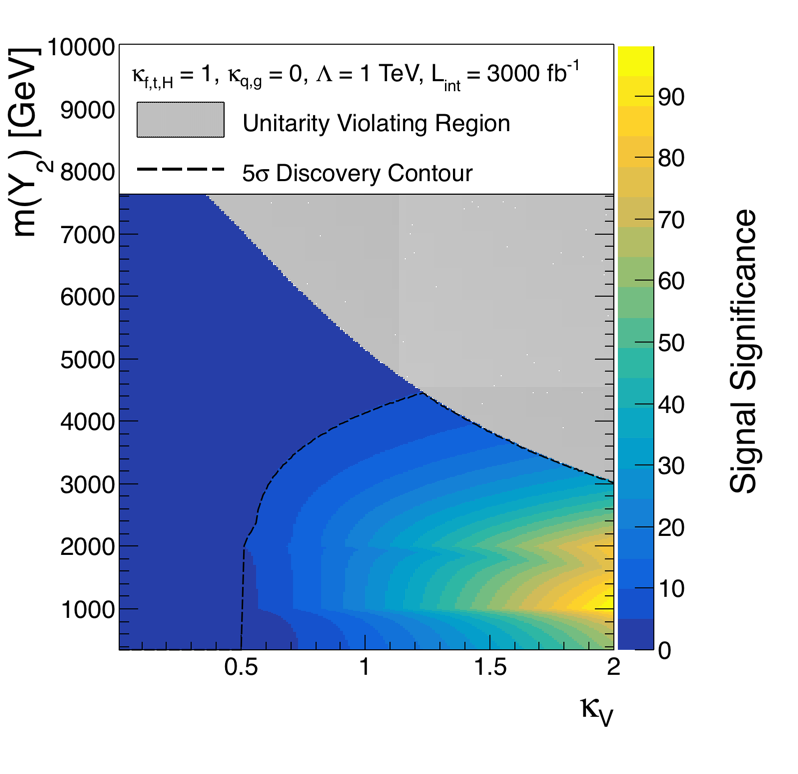}
 \end{center}
 \caption{Expected signal significance for the proposed VBF final state. The results are shown as $m(Y_{2})$ vs. $\kappa_{V}$ on the $y-x$ plane, and the expected signal significance on the $z-$axis.  The expected signal significance was calculated performing a profile binned likelihood of the diphoton mass distribution, assuming an expected luminosity of 3000 $fb^{-1}$. The black dashed line encloses the region with 5$\sigma$ discovery potential.}
 \label{fig:resultsb}
 \end{figure}

The expected signal significance has been calculated considering different scenarios of integrated luminosity at the LHC, ranging from 100 to 3000 fb$^{-1}$ (based on the current and projected luminosity at the LHC), 
and assuming different values of $\kappa_{V}$. The results are  presented in Figs.~\ref{fig:results} and~\ref{fig:resultsb}, as $m(Y_{2})$ vs. $\kappa_{V}$ on the $y-x$ plane, and the expected signal significance (using the shape based approach) on the $z-$axis. Similar to Figs.~\ref{fig:UVphasespaceA}-~\ref{fig:UVphasespaceB}, only the unitarity acceptable region $\Gamma < m(Y_{2})$ in the EFT is considered for the calculation of the signal significances. 
The black dashed line in Figs.~\ref{fig:results} and~\ref{fig:resultsb} delimits the 5$\sigma$ discovery contour. Assuming an integrated luminosity of 100 fb$^{-1}$ (3000 fb$^{-1}$), there is discovery potential for $\kappa_{V} > 1.85$ (1.7) in the entire range of theoretically allowed values for $m(Y_{2})$ (up to 4.4 TeV). For lower couplings of $\kappa_{V} = 0.6$ (1.0), the proposed VBF $Y_{2}\to\gamma\gamma$ search can provide signal significances greater than 5$\sigma$ for $Y_{2}$ masses up to 3.0 (4.2) TeV with the amount of pp data expected from the high-luminosity LHC. 

\section{Discussion}

The main result of this paper is that probing heavy spin-2 bosons $Y_{2}$ produced through VBF processes can be a key methodology to complement current searches for TeV scale particles at the LHC. In the context of a generic and model independent approach via an effective field theory, both standard and non-universal couplings for $Y_{2}$ to the SM particles can be studied.  Of particular interest is that there has been no evidence for a TeV scale $Y_{2}$ at the LHC to date.  The $Y_{2}$ boson may have failed to be discovered in searches within the CMS and ATLAS experiments thus far because $\kappa_{q,g}$ is small.  Thus, the $Y_{2}$ would not be discoverable by the ``standard'' searches via gluon-gluon fusion or quark-antiquark annihilation currently conducted at the LHC.  In addition, even if a $Y_{2}$ were to be discovered with $gg\to{Y_{2}}$ or $q\bar{q}\to{Y_{2}}$, a VBF $Y_{2}$ search remains a key part of the $Y_{2}$ search program at the LHC in order to establish the couplings $\kappa_{V}$ of the $Y_{2}$ to the SM vector bosons.  We consider $Y_{2}$ decays to $\gamma\gamma$ to show that the requirement of a diphoton pair combined with two high $p_{T}$ forward jets with large separation in pseudorapidity and with large dijet mass is effective in reducing contributions from QCD multijet, $\gamma$+jets, $\gamma\gamma$+jets, and other SM backgrounds. Dilepton, dijet, and $ZZ$/$WW$ channels can all result from spin-1 $Z'$ or spin-2 $Y_{2}$ production.  However, an excess in the $\gamma\gamma$ final state would rule out $Z'$.  In keeping within the EFT in this analysis, various values of $\kappa_{V}$ have been studied, and effects of varying $\kappa_{V}$ on the branching ratio and decay width for $Y_{2}$ have been considered.  These considerations related to $\kappa_{V}$ have been made with particular attention to the theoretically allowed phase space for determining discovery potential.  Assuming proton-proton collisions at $\sqrt{s} = 13$ TeV, the proposed VBF $Y_{2}\to\gamma\gamma$ search is expected to achieve a discovery reach with signal significance greater than 5$\sigma$ for $Y_{2}$ masses up to 4.4 TeV and $\kappa_{V}$ couplings down to 0.5.
  
\section{Acknowledgements}

We thank the constant and enduring financial support received for this project from the faculty of science at Universidad de los Andes (Bogot\'a, Colombia), the administrative department of science, technology and innovation of Colombia (COLCIENCIAS), the Physics \& Astronomy department at Vanderbilt University and the US National Science Foundation. This work is supported in part by NSF Award PHY-1806612.

\end{document}